\documentclass[aps,prb,superscriptaddress,reprint]{revtex4-1}
\usepackage{graphicx}

\usepackage{dcolumn}

\begin{document}


\title{Electronic transitions of single silicon vacancy centers in the near-infrared spectral region}


\author{Elke Neu}
\affiliation{Universit\"at des Saarlandes, Fachrichtung 7.2 (Experimentalphysik), 66123 Saarbr\"ucken, Germany}
\author{Roland Albrecht}
\affiliation{Universit\"at des Saarlandes, Fachrichtung 7.2 (Experimentalphysik), 66123 Saarbr\"ucken, Germany}
\author{Martin Fischer}
\affiliation{Universit\"at Augsburg, Lehrstuhl f\"ur Experimentalphysik IV, 86135 Augsburg, Germany}
\author{Stefan Gsell}
\affiliation{Universit\"at Augsburg, Lehrstuhl f\"ur Experimentalphysik IV, 86135 Augsburg, Germany}
\author{Matthias Schreck}
\email[]{matthias.schreck@physik.uni-augsburg.de}
\affiliation{Universit\"at Augsburg, Lehrstuhl f\"ur Experimentalphysik IV, 86135 Augsburg, Germany}
\author{Christoph Becher}
\email[]{christoph.becher@physik.uni-saarland.de}
\affiliation{Universit\"at des Saarlandes, Fachrichtung 7.2 (Experimentalphysik), 66123 Saarbr\"ucken, Germany}


\date{\today}

\begin{abstract}
Photoluminescence (PL) spectra of single silicon vacancy (SiV) centers in diamond frequently feature very narrow room temperature PL lines in the near-infrared (NIR) spectral region, mostly between \mbox{820 nm} and  \mbox{840 nm}, in addition to the well known zero-phonon-line (ZPL) at approx.\ 738 nm [E. Neu et al., Phys.\ Rev.\ B \textbf{84}, 205211 (2011)].  We here exemplarily prove for a single SiV center that this NIR PL is due to an additional purely electronic transition (ZPL). For the NIR line at 822.7 nm, we find a room temperature linewidth of 1.4 nm (2.6 meV). The line saturates at similar excitation power as the ZPL. ZPL and NIR line exhibit identical polarization properties. Cross-correlation measurements between the ZPL and the NIR line reveal anti-correlated emission and prove that the lines originate from a single SiV center, furthermore indicating a fast switching between the transitions (0.7 ns). $g^{(2)}$ auto-correlation measurements exclude that the NIR line is a vibronic sideband or that it arises due to a transition from/to a meta-stable (shelving) state.
\end{abstract}

\pacs{}

\maketitle

\section{Introduction}\label{Ir PL}
In recent years, color centers in diamond have been investigated as solid state light emitters especially in the context of single photon sources (e.g., Refs.\ \onlinecite{Brouri2000,Kurtsiefer2000,Wang2006,Neu2011,Gaebel2004}) or, if incorporated into nanodiamonds, as fluorescence labels for \textit{in vivo} imaging (e.g., Refs.\ \onlinecite{Chang2008,Mohan2010,Neugart2007}). For these applications, color centers in diamond stand out owing to feasible room temperature operation as well as a high photostability. For applications in \textit{in vivo} imaging, biocompatibility \cite{Mohan2010} of the nanodiamonds and a large variety of possible surface functionalizations \cite{Neugart2007} add to the advantages. The majority of previous experiments utilizes the well known nitrogen vacancy (NV) color center in diamond. Nevertheless, NV centers suffer from a major disadvantage, namely their broad room temperature emission bandwidth of about \mbox{100 nm} induced by strong electron-phonon coupling.\cite{Kurtsiefer2000} Recently, silicon vacancy (SiV) centers emerged as a promising alternative.\cite{Wang2006,Neu2011,Neu2011a,Neu2011b} In contrast to NV centers, SiV centers enable narrow bandwidth room temperature single photon emission with a linewidth down to \mbox{0.7 nm} and a high brightness (up to \mbox{5 Mcps}).\cite{Neu2011} The narrow bandwidth results from low electron-phonon coupling leading to a significant concentration (exceeding 70\%) of the luminescence in the narrow, purely electronic transition, i.e., the zero-phonon-line (ZPL), at approx.\ \mbox{738 nm}. Furthermore, single SiV centers enable fully linearly polarized single photon emission,\cite{Neu2011b} which is advantageous for applications in quantum cryptography \cite{Bennett1984} or in the frequency conversion of single photons.\cite{Zaske2011} SiV centers can be efficiently produced \textit{in situ} by incorporation of Si impurities during  chemical vapor deposition (CVD) of diamond.\cite{Barjon2005,Dollinger1995} Furthermore, the production of colloidal solutions of fluorescent nanodiamonds containing SiV centers from polycrystalline CVD diamond films has been demonstrated.\cite{Neu2011a} Thus, SiV centers are also promising candidates for fluorescence labels especially due to feasible excitation with red laser light as well as their narrow emission in the red/near-infrared spectral region.\cite{Chang2008,Weissleder2003} For single photon emission as well as fluorescence labels, a further extension of color center emission into the longer wavelength near-infrared spectral range is of interest.
Despite the promising applications introduced above, several fundamental characteristics of the SiV complex are still unsettled: The charge state of the SiV center responsible for the \mbox{738 nm} ZPL as well as the spatial symmetry of the complex are still under debate.\cite{Moliver2003,Goss1996}  In addition, a generally accepted suggestion for the electronic level scheme of the SiV center is still lacking.

In this work, we aim at further elucidating the electronic level scheme as well as the possible electronic transitions of single SiV centers. In Ref.\ \onlinecite{Neu2011b}, we found that
the photoluminescence (PL) spectra of single SiV centers frequently feature very narrow room temperature PL lines in the near-infrared spectral region, mostly between \mbox{820 nm} and  \mbox{840 nm}. The present work is dedicated to an extensive investigation of these lines which we will term \textit{near-infrared (NIR) line(s)}. The NIR lines might arise from narrow vibronic sidebands or due to additional purely electronic transitions of the SiV centers. In previous work, Sittas et al.\cite{Sittas1996} found a significant spectral narrowing upon cooling for three spectral features (\mbox{776 nm}, \mbox{797 nm} and \mbox{812 nm}) in the sideband region of SiV center ensemble PL. Thus, they attributed these features to purely electronic rather than vibronic transitions. In earlier investigations, a three level model for single SiV centers has been established (see Refs.\ \onlinecite{Neu2011,Wang2006}). It includes an excited and ground state giving rise to the ZPL transition and a third meta-stable level, termed shelving state. In the framework of this model, one might suspect that the NIR lines are due to a fluorescent transition from the excited state to the shelving state or due to a transition from the shelving state to the ground state. In Ref.\ \onlinecite{Dhaenens2011}, absorption measurements of single crystal CVD diamond samples containing high densities of SiV centers are reported: The authors found absorption lines in a wavelength range (830--860 nm) similar to the wavelength range of the NIR PL we observe here. The detection of a transition in absorption indicates that it involves the ground state of the system and only for purely electronic transitions, absorption and emission occur at the same wavelength. One thus might suspect that the NIR PL is attributed to an electronic transition from the shelving state to the ground state. Similarly, for NV centers, radiative transitions between shelving states (singlet states) have been reported at \mbox{1046 nm}.\cite{Rogers2008} However, in the context of single color center spectroscopy, one has to make sure that the sharp PL lines do not belong to other color centers incorporated together with the SiV centers during CVD growth (e.g., nickel-related centers emitting between \mbox{806 nm} and \mbox{820 nm}).\cite{Aharonovich2008} In this work, we prove exemplarily for a single SiV center with pronounced NIR emission that the NIR line arises from an additional, independent purely electronic transition that cannot be explained within the established three level model. Section \ref{sec:samples} introduces the investigated sample as well as the experimental setup. In Sec.\ \ref{sec:spectr}, we discuss the spectrum of the SiV PL at room temperature. In Sec.\ \ref{sec:satmeasIR}, we investigate the saturation curves of the ZPL and NIR PL. The polarization properties will be addressed in Sec.\ \ref{sec:polmeas}. In Sec.\ \ref{sec:IrPLcorr}, we present intensity auto-correlation $g^{(2)}$ measurements of the ZPL and the additional NIR line as well as cross-correlation measurements of  both lines. The measurements prove single photon emission and reveal the population dynamics of the SiV center. The cross-correlation measurements prove that the NIR line originates from the same color center as the well known ZPL. In Sec.\ \ref{sec:discussg2}, we discuss the measurements and identify the NIR line as an independent electronic transition not connected to the shelving state.
\section{Sample preparation and experimental setup \label{sec:samples}}
The investigated SiV centers are contained in heteroepitaxial diamond films synthesized by microwave plasma chemical vapor deposition (MPCVD) on Ir/YSZ/Si(001) substrates. The preparation of these substrates comprises the pulsed laser deposition (PLD) of an yttria-stabilized zirconia (YSZ) buffer layer followed by the e-beam evaporation of a single crystal Ir film on a 4$^{\circ}$ off-axis Si(001) wafer (for further details see Ref.\ \onlinecite{Gsell2009}). The mosaic spread of the Ir film is 0.15--0.2$^{\circ}$. For the roughness we measure a value of $\le$1 nm RMS.
The bias enhanced nucleation (BEN) procedure is applied in a MPCVD reactor in order to generate heteroepitaxial diamond nuclei. During BEN the gas mixture in the reactor consists of 3\% CH$_4$ in H$_2$ at a pressure of 40 mbar. A plasma discharge is ignited by feeding in 2000 W of microwave power. A negative bias voltage of $-300$ V is applied to the Ir/YSZ/Si(001) substrate. After the BEN step, growth is performed without bias for 20 min at a substrate temperature of 800 $^{\circ}$C and a reduced methane concentration of 0.5\%. Scanning electron micrographs show a closed diamond layer. Its thickness is 90 nm as determined by ellipsometry measurements. SiV centers are created \textit{in situ} with low density as a result of plasma etching of the Si substrates and subsequent incorporation of Si into the growing diamond.\cite{Neu2011}

We investigate single SiV centers using confocal laser fluorescence microscopy. Optical excitation of the color centers is performed using a cw frequency-doubled, diode-pumped solid state laser providing an excitation wavelength of \mbox{671 nm} corresponding to a photon energy of \mbox{1.85 eV}. The excitation wavelength coincides with absorption bands of the SiV center.\cite{Iakoubovskii2001} Furthermore, it avoids direct photoionization of the SiV centers as the optical ionization threshold, i.e., the energetic distance of the ground state of the SiV center to the conduction band edge, is  \mbox{2.05 eV} (\mbox{605 nm}).\cite{Iakoubovskii2000b}  The laser is focussed onto the sample by a high numerical aperture microscope objective (Olympus, magnification 100x, NA 0.8). The fluorescence is collected using the same objective and separated from reflected laser light by a dichroic mirror and dielectric longpass filters. For correlation measurements, we employ a Hanbury Brown Twiss (HBT) setup with two avalanche photodiodes (APDs, Perkin Elmer SPCM AQRH-14) featuring a typical quantum efficiency of approx.\ 65\% at \mbox{740 nm} and  50\% at \mbox{820 nm}. Dielectric bandpass filters can be inserted in front of each APD to select the investigated luminescence lines (\mbox{730--750 nm} for the ZPL, \mbox{815--825 nm} for the NIR line). Using this experimental configuration enables the simultaneous measurement of saturation curves for the ZPL and the NIR line as well as intensity cross-correlation measurements. Photon arrival times are recorded with a fast timing electronics (Pico Quant, Pico Harp, timing resolution of electronics 4 ps, APD timing jitter 354 ps). We use this data to calculate the correlation functions. To investigate the polarized absorption of single SiV centers, the (linear) excitation polarization is rotated using a half-wave plate. The polarization of the emitted PL is investigated using a linear polarization analyzer. To investigate the spectra of the color center PL, we use a grating spectrometer (Horiba Jobin Yvon, iHr550). A grating with  600 grooves/mm enables a resolution of approx.\ 0.18 nm. All experiments were performed at room temperature.
\section{Spectroscopic characterization \label{sec:spectr}}
\begin{figure}[h]
\centering
\includegraphics{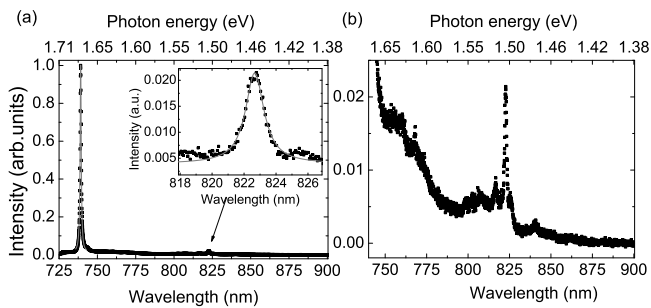}
\caption{PL spectrum of a single SiV center. All spectra have been normalized to the ZPL. (a) Overview spectrum revealing a bright ZPL with low sideband contributions. The solid lines give Lorentzian fits to the data. The inset shows the dominant narrow PL line in the region $\lambda>$ 800 nm. (b) Zoomed image of the spectral region revealing the vibronic sidebands. \label{fig:specSiV}}
\end{figure}

Figure \ref{fig:specSiV} displays the spectrum of a single SiV center located in the (001) heteroepitaxial diamond film described in Sec.\ \ref{sec:samples}. The ZPL is clearly visible at \mbox{739.1 nm}, a Lorentzian fit yields a linewidth of \mbox{0.9 nm} (\mbox{2.0 meV}). The peak wavelength and the width of the ZPL is in accordance with previous observations for single SiV centers in randomly oriented nanodiamonds as well as heteroepitaxial nanoislands.\cite{Neu2011,Neu2011b} We point out that we observe a considerable inhomogeneous spread of the ZPL wavelengths of individual SiV centers in the investigated heteroepitaxial diamond film in accordance with our previous investigations:\cite{Neu2011b} During the course of our investigations, single emitters with wavelengths ranging from 735.3--739.2 nm have been observed. The inhomogeneous spread of ZPL peak positions is related to micro-stress fields at the locations of the individual color centers. Different stress sources are crucial in heteroepitaxial diamond, e.g., stress due to thermal expansion mismatch of substrate and diamond as well as growth stress developing between merging diamond grains. Since the relationship between the energetic shift of the SiV center ZPL and the local stress state is not known in detail, we cannot use the measured shifts to estimate the stress amplitude (for a detailed discussion see Ref.\ \onlinecite{Neu2011b}).

Figure \ref{fig:specSiV}(b) shows the sideband region of the spectrum in detail. As apparent from Fig.\ \ref{fig:specSiV}(b), a single line dominates the spectrum in the wavelength range $\lambda>\mbox{800 nm}$. A Lorentzian fit to the narrow NIR line yields a peak position of \mbox{822.7 nm} and a linewidth of \mbox{1.4 nm} (\mbox{2.6 meV}) [see inset of Fig.\ \ref{fig:specSiV}(a)]. Thus, the NIR line has a linewidth comparable to the ZPL. The energy difference between the ZPL and the NIR line is \mbox{170 meV}, close to the energy of the Raman phonon mode in diamond (\mbox{165 meV}, see, e.g., Ref.\ \onlinecite{Solin1970}). However, due to the different spectral widths as well as the results of the $g^{(2)}$ measurements discussed below, we exclude that the \mbox{822.7 nm} NIR line originates from Raman scattering of the \mbox{739.1 nm} ZPL. Furthermore, the comparably high intensity of the NIR line excludes Raman scattered light. For single SiV centers (see Ref.\ \onlinecite{Neu2011b}), PL spectra often feature multiple NIR lines with wavelengths in the range from \mbox{820 nm} to \mbox{840 nm}, additionally excluding Raman scattered light. In low temperature experiments, we find a spectral narrowing of NIR lines down to 0.3 nm (0.56 meV) at 5 K accompanied by a blue shift of 1.6 nm. In accordance with Ref.\ \onlinecite{Sittas1996}, this observation supports the identification of the NIR lines as electronic transitions.

The spectrum given in Fig.\ \ref{fig:specSiV} does not straightforwardly allow for a comparison of the relative intensities of the emission lines as the detection efficiency varies over the observed wavelength range. Using the manufacturer supplied efficiency data for the spectrometer grating as well as the employed CCD detector, we find that the efficiency at 820 nm amounts to roughly 75\% of the efficiency at 740 nm. Furthermore, the confocal setup has been optimized for the detection of fluorescence at a wavelength of 740 nm, thus other wavelengths might not be optimally mapped. A more precise determination and comparison of the brightness of the emission lines is performed in Sec.\ \ref{sec:satmeasIR} using the single photon rates obtained with the APDs of the HBT setup. With this method, the setup can be individually optimized for the different emission lines.
\section{Saturation measurements \label{sec:satmeasIR}}
\begin{figure}[h]
\centering
\includegraphics{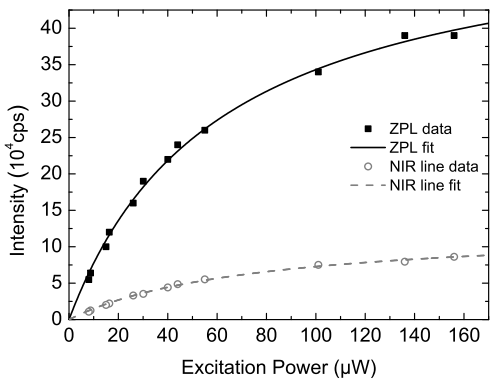}
\caption{Saturation curves measured for the single SiV center. Filled squares (open circles) represent the datapoints, the solid line (dashed line) represents the fit using Eq.\ (\ref{Eq:satfunc}) for the ZPL (NIR line). The measurement for the NIR line has been corrected for the lower detection efficiency.  \label{fig:satSiV}}
\end{figure}

As apparent from Fig.\ \ref{fig:specSiV}(a), the \mbox{822.7 nm} line is significantly weaker than the ZPL at \mbox{739.1 nm}. To quantify the PL intensity ratio, we perform a simultaneous measurement of the excitation power dependent photon count rates $I(P)$ on these lines as described in Sec.\ \ref{sec:samples}. Figure \ref{fig:satSiV} displays the saturation curves measured on the ZPL and the \mbox{822.7 nm} line. The fluorescence count rate $I(P)$ obtained for a single color center is described by
\begin{equation}
      I(P)=I_\infty\frac{P}{P+P_{sat}}
      \label{Eq:satfunc}
\end{equation} if no linearly rising background luminescence is present. $P_{sat}$ is the saturation power, $I_{\infty}$ the maximum obtainable photon rate.  Fits using Eq.\ (\ref{Eq:satfunc}) describe the measured data very well (see Fig.\ \ref{fig:satSiV}), indicating a negligible contribution of background luminescence. For the ZPL, we find $P_{sat}=61\pm4$ $\mu$W and $I_{\infty}=(55\pm2)\times10^4$ cps. For the \mbox{822.7 nm} line, we find $P_{sat}=75\pm4$ $\mu$W and $I_{\infty}=(12.8\pm0.4)\times10^4$ cps. The very similar saturation powers obtained for the two lines can be considered as first evidence that the PL is excited via the same absorptive transition. This in turn indicates that the lines stem from the same color center. Note that the count rate for the NIR line has been corrected for the lower detection efficiency of the APD at \mbox{820 nm} as well as a lower bandpass filter transmission. Comparing the photon count rates, the ZPL intensity amounts to 4.3 times the intensity of the \mbox{822.7 nm} line (the non-corrected count rate of the ZPL is about an order of magnitude higher than the NIR count rate).
We point out that the NIR as well as ZPL emission were perfectly photostable: We did not detect any fluorescence intermittence during the course of measurements presented in this paper (checked for time intervals down to 1 ms). The maximum excitation power used amounts to three times the saturation power $P_{sat}$, proving photostability also for saturated excitation.
\section{Polarization measurements \label{sec:polmeas}}
Further information about luminescent transitions can be obtained using polarization spectroscopy.\cite{Kaplyanskii1963}  First, we measure the polarized absorption of the SiV center by rotating the excitation polarization and simultaneously detecting the PL either on the \mbox{739.1 nm} ZPL or on the  \mbox{822.7 nm} line. Second, we measure the linear polarization degree of the emitted light.

\begin{figure}[h]
\centering
\includegraphics{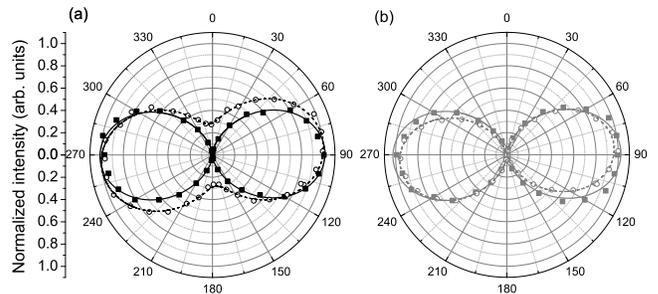}
\caption{Polarization measurements for a single SiV center: (a) measurements for the \mbox{822.7 nm} line, (b) measurements for the \mbox{739.1 nm} ZPL. Filled squares (open circles) represent the measured data of polarized absorption (emission polarization). Solid lines (dashed lines) give the sinusoidal fits to the absorption (emission) data. To enable comparison of the data, all measurements have been normalized to one.  \label{fig:pol740820} }
\end{figure}

Figure \ref{fig:pol740820}(a) [Fig.\ \ref{fig:pol740820}(b)] displays the measurements for the \mbox{822.7 nm} [\mbox{739.1 nm}] line in a polar plot (details see figure caption). The polarized absorption, i.e., the polarization dependence of the excitation efficiency, indicates the preferential absorption of linearly polarized light, consistent with a single absorption dipole, with a polarization direction of 89$^{\circ}$ for ZPL and NIR line. For light perpendicular to that direction, effectively no absorption takes place. We emphasize that polarized absorption only addresses the dipole component in the sample plane, i.e., the plane perpendicular to the excitation laser propagation direction.\cite{Ha1999} Thus, the two lines arise from an absorptive transition with the same in plane dipole moment. Note that an orientation of 0 or 90$^{\circ}$ corresponds to an orientation along the $\langle110\rangle$ crystal directions in the (001) plane. Thus the observed orientation of the present individual SiV center is in accordance with our previous observations in Ref.\ \onlinecite{Neu2011b} which were obtained using the ZPL emission. Together with the observation of a similar saturation power, this result supports the interpretation that the lines are excited via the same absorptive transition and thus arise from the same emitter.

The PL shares a common linear polarization direction within the experimental error of approx. $\pm1^{\circ}$ (84.0$^{\circ}$ for the \mbox{822.7 nm} line and 85.6$^{\circ}$ for the \mbox{739.1 nm} ZPL). Figure\ \ref{fig:pol740820}(b) shows a visibility close to 100\%, while Fig.\ \ref{fig:pol740820}(a) suggests a lower visibility. However, the reduced contrast is due to technical reasons, i.e., a reduced performance of the employed polarization analyzer in the NIR spectral range. Background fluorescence has been subtracted for both  measurements; however, due to a spatially fluctuating background in the vicinity of the color center, this procedure may introduce an error in the visibility of 5--10\%. The observation of a common polarization direction also indicates a common orientation of the emission dipoles in the sample plane. This result might be a hint that the lines originate from the same emitter.  For the ZPL  and the NIR line, polarized absorption and emission yield an almost parallel orientation of the emission and absorption dipoles, respectively, (see Fig.\ \ref{fig:pol740820}) in accordance with previous findings for the ZPL of single SiV centers.\cite{Neu2011b}

The three-dimensional dipole orientation can only be retrieved using statistics for a large number of emitters that reveal the different equivalent directions (see, e.g., Ref.\ \onlinecite{Neu2011b}). The investigation of the three-dimensional dipole orientation for the NIR transition is beyond the scope of this work.
\section{Intensity correlation measurements \label{sec:IrPLcorr}}
\begin{figure}[h]
\includegraphics{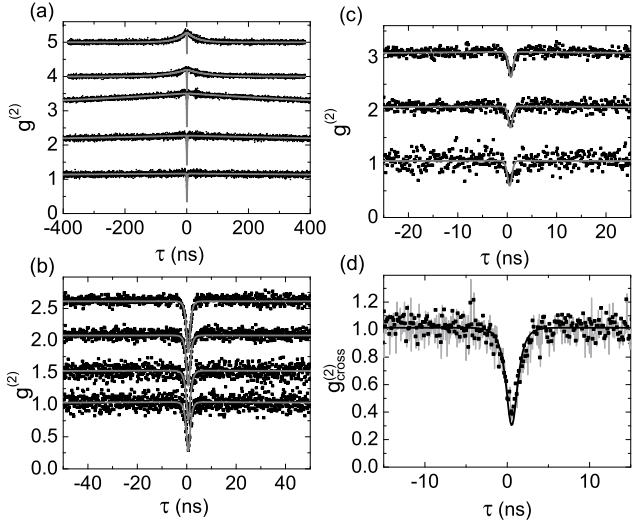}
\caption{Intensity correlation measurements (normalized assuming $g^{(2)}(\tau)=1$ for long delay times $\tau$). Note that adjacent $g^{(2)}$ functions have been shifted for clarity [(a)+(c): adjacent $g^{(2)}$ functions shifted by 1, (b):  adjacent $g^{(2)}$ functions shifted by 0.5]:  (a)  $g^{(2)}_{ZPL}$ (739.1 nm line)  $0.62\,P_{sat}-2.92\,P_{sat}$, (b) $g^{(2)}_{ZPL}$ $0.08\,P_{sat}-0.37\,P_{sat}$, (c)  $g^{(2)}_{NIR}$ (822.7 nm line) with increasing excitation power ($0.32\,P_{sat}$, $0.81\,P_{sat}$, $1.49\,P_{sat}$), (d) intensity cross-correlation $g^{(2)}_{cross}$ of 739.1 nm  and 822.7 nm line (approx. $0.3\,P_{sat}$). (further explanation see text). Solid lines represent fits using Eq.\ (\ref{g23level}) convoluted with the instrument response function of the HBT setup.  \label{fig:g2}}
\end{figure}

In this section, we investigate the luminescent transitions by measuring the intensity auto-correlation $g^{(2)}$  as well as the cross-correlation $g^{(2)}_{cross}$ of the PL lines. $g^{(2)}$ measurements can be used to prove the single photon nature of the PL as well as to analyze the population dynamics of the color center under investigation.\cite{Neu2011,Aharonovich2010a,Wu2006} $g^{(2)}_{cross}$ measurements can prove that multiple fluorescence lines originate from the same emitter as demonstrated in Ref.\ \onlinecite{Gaebel2006b} for the PL from different charge states of an NV center.
Figure \ref{fig:g2} summarizes the excitation power dependent intensity correlation measurements performed for the single SiV center.

Figure \ref{fig:g2}(a)+(b) displays the $g^{(2)}$ measurements for the \mbox{739.1 nm} ZPL ($g^{(2)}_{ZPL}$) revealing a distinct antibunching close to  zero delay. For intermediate delays $\tau$, $g^{(2)}_{ZPL}$  exceeds one (bunching). To describe the intensity dependent $g^{(2)}_{ZPL}$ functions, we use an extended three level scheme depicted in Fig.\ \ref{5levelscheme}. The transition $2 \to 1$ marks the ZPL. If the color center resides in state 3 (shelving state), no photons on the ZPL are emitted.\cite{Neu2011} To obtain the $g^{(2)}$ function, one solves the rate equations for the populations $N_i$:
\begin{eqnarray}
\frac{dN_1}{dt}&=&N_2k_{21}-N_1k_{12}+N_3k_{31}\label{ratengl1}\\
\frac{dN_2}{dt}&=&-N_2k_{21}+N_1k_{12}-N_2k_{23}\label{ratengl2}\\
\frac{dN_3}{dt}&=&N_2k_{23}-N_3k_{31}\label{ratengl3}
\end{eqnarray}
We assume that the system is in the ground state at time zero and that the sum of the populations equals one.
$g^{(2)}_{ZPL}$ is given by $\frac{N_2(\tau)}{N_2(\tau\rightarrow \infty)}$ (Ref.\ \onlinecite{Kurtsiefer2000}) resulting in
\begin{equation}
g^{(2)}_{ZPL}(\tau)=1-(1+a)\,e^{-|\tau|/\tau_1}+a\,e^{-|\tau|/\tau_2}
\label{g23level}
\end{equation}
The parameters $a$, $\tau_1$ and $\tau_2$ are given by:\cite{Neu2011}
\begin{eqnarray}
            \label{taupar1}\tau_{1,2}=&2/(A\pm\sqrt{A^2-4B})\\[0.01 \textwidth]
            \label{taupar2}A=&k_{12}+k_{21}+k_{23}+k_{31}\\[0.01 \textwidth]
            \label{taupar3}B=&k_{12}k_{23}+k_{12}k_{31}+k_{21}k_{31}+k_{23}k_{31}\\[0.01 \textwidth]
            \label{apar} a=&\frac{1-\tau_2k_{31}}{k_{31}(\tau_2-\tau_1)}
\end{eqnarray}
The parameter $\tau_1$ governs the antibunching for short delays $\tau$, while the parameter $\tau_2$ governs the bunching at intermediate time\-scales. The parameter $a$ determines how pronounced the bunching is. To describe the excitation power dependence of the parameters $a$, $\tau_1$ and $\tau_2$, we use the model introduced in Ref.\ \onlinecite{Neu2011}, where a linear dependence of the pumping rate coefficient $k_{12}$ of the excitation power $P$ is assumed ($k_{12}=\sigma P$). Furthermore,  the color center can be re-excited from the shelving state via a fourth state (see Fig.\ \ref{5levelscheme}) leading to a power dependent de-shelving rate coefficient $k_{31}$ with  \begin{equation}
k_{31}=\frac{d\cdot P}{P+c}+k_{31}^0, \label{satdeshrate}
\end{equation}
where $k_{31}^0$ is the power independent de-shelving coefficient and $c$ and $d$ describe the power dependence of $k_{31}$. In this model, all rate parameters can be determined using limiting values of the parameters $a$, $\tau_1$ and $\tau_2$ at high and low power as well as a fit of the power dependent curves of these parameters (for details see Ref.\ \onlinecite{Neu2011}).

\begin{figure}[h]
\centering
\includegraphics{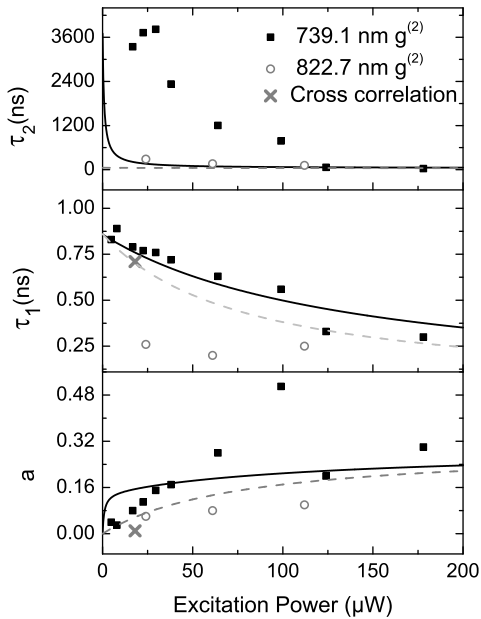}
\caption{Parameters  $a$, $\tau_1$ and $\tau_2$ obtained for the $g^{(2)}$ measurements. Black solid lines give the intensity dependent curves for the ZPL $g^{(2)}$ function in the intensity dependent de-shelving model. Gray, dashed curves are plotted using the model with constant de-shelving (explanation see text).   \label{fig:g2par}}
\end{figure}

We use Eq.\ (\ref{g23level}) convoluted with the instrument response function of our HBT setup to fit the measured data. The instrument response function of the setup is obtained independently via measuring the intensity auto-correlation function of attenuated ultrafast laser pulses from a mode-locked titanium-sapphire laser (Tsunami Spectra Physics, nominal pulse duration 100 fs). The instrument response is well described by a Gaussian function with a full width at half maximum of 830 ps. The analytical result of the convolution of Eq.\ (\ref{g23level}) with the instrument response is used as fitting function. The fitted functions are fully consistent with the measured data, thus proving single photon emission with negligible background contributions. The fits provide the parameters $a$, $\tau_1$ and $\tau_2$, summarized  in Fig.\ \ref{fig:g2par}. We deduce the parameters describing the color center in the above specified model with intensity dependent de-shelving: $k_{23}= 5.8$ MHz, $k_{21}=1.16$ GHz, $k_{31}^0=0.27$ MHz, $d=18.96$ MHz, $\sigma=8.4$ MHz/$\mu$W and $c=66.6$ $\mu$W. Thus, the rate coefficient for the ZPL transition ($k_{21}$) is the largest coefficient, while the rate coefficient populating the shelving state ($k_{23}$) is significantly smaller. The de-shelving rate coefficient at low power  ($k_{31}^0$) is nearly an order of magnitude smaller than its high power limit ($d+k_{31}^0$).  Figure \ref{fig:g2par} also displays the power dependent curves for $\tau_1$, $\tau_2$ and $a$  obtained from this model as solid lines and the curves from a model with constant $k_{31}$ as dashed lines (for details see Ref.\ \onlinecite{Neu2011}). As apparent from Fig.\ \ref{fig:g2par}, the increase of $\tau_2$ at low excitation powers is overestimated by the intensity dependent de-shelving model, whereas it is not at all described using the model with constant de-shelving.  The power dependence of $\tau_1$ and $a$ is reasonably described in the model including intensity dependent de-shelving, fitting the data more closely than the model with constant $k_{31}$. These observations thus support the assumption of an intensity dependent de-shelving process.

Figure \ref{fig:g2}(c) displays the $g^{(2)}$ auto-correlation measurements of the \mbox{822.7 nm} fluorescence ($g^{(2)}_{NIR}$). Fewer measurements have been performed as the measurement times are significantly increased due to the low count rates. The fits of $g^{(2)}_{NIR}$ in Fig.\ \ref{fig:g2}(c) show that the deviation from $g^{(2)}(0)=0$ is fully explained by the instrument response of our HBT setup which has been taken into account for the fits.  As apparent from Figs.\ \ref{fig:g2}(c) and \ref{fig:g2par}, $g^{(2)}_{NIR}$ displays only a very weak bunching. Fitting these $g^{(2)}$ functions analogously to the ones of the ZPL, the antibunching time constants $\tau_1^{NIR}$ are about a factor of 3 shorter than for $g^{(2)}_{ZPL}$  (see Fig.\ \ref{fig:g2par}, low power limit $\tau_1^{NIR,0}=0.26$ ns, $\tau_1^{0}=0.86$ ns).

In addition to the $g^{(2)}$ auto-correlation measurements, we perform a $g^{(2)}_{cross}$ cross-correla\-tion measurement of the ZPL and NIR emission lines. If these lines originate from the same emitting color center, a $g^{(2)}_{cross}$ measurement reveals an antibunching: the emission is anti-correlated. If the lines originate from different emitters, one expects no correlation between the fluorescence photons.\cite{Gaebel2006b} We point out that the $g^{(2)}_{cross}$ measurement does not enable to discriminate between lines from different charge states of the emitter or lines from different transitions (purely electronic or vibronic) of the same charge state.

$g^{(2)}_{cross}$ measurements have been used to prove the charge state conversion of a single NV center from NV$^-$ to NV$^0$.\cite{Gaebel2006b} We here assume that the \mbox{822.7 nm} line is not due to a second charge state, as SiV$^0$ emission  has been reported at \mbox{946 nm}.\cite{Dhaenens2011} Nevertheless, also the improbable situation of a single electron alternatively charging an SiV center (emitting at \mbox{739.1 nm}) and another color center (emitting at \mbox{822.7 nm}) would lead to anti-correlated emission. However, one might also expect to observe fluorescence blinking in this case, which we do  not find in our experiments. Additionally, the similar saturation curves and polarization measurements support the assumption that the lines arise from a single emitter.  Figure \ref{fig:g2}(d) displays the measurement of  $g^{(2)}_{cross}$. $g^{(2)}_{cross}$ has been obtained using two methods. First, we calculate the cross-correlation by correlating each time tag (photon arrival time) recorded on one avalanche photodiode with all other time tags for the second diode [filled squares in Fig.\ \ref{fig:g2}(d)] analogously to the auto-correlation measurements. Second, we simulate a start-stop measurement by using each detection event on the NIR line as a start event and search for the next emission of a ZPL photon which serves as the stop event and determines the delay $\tau$ [light gray line in Fig.\ \ref{fig:g2}(d)]. The second method has previously been used to obtain cross-correlation functions for quantum dot emission lines.\cite{Kiraz2002,Regelman2001}  We find strong antibunching, the curves calculated using the two methods are equal within experimental errors. The absence of an asymmetry in the start-stop measurement proves that there is no preferred time-ordering of the emission events of both lines (see also discussion Sec.\ \ref{sec:discussg2}).  The time constant $\tau_1^{cross}=\mbox{0.7 ns}$  is close to the values obtained for $g^{(2)}_{ZPL}$ [fit: black line in Fig.\ \ref{fig:g2}(d)]. In summary, the $g^{(2)}_{cross}$ measurement identifies the \mbox{822.7 nm} PL as anti-correlated to the SiV ZPL. Thus, we conclude that the lines originate from the same emitter.
\section{Discussion of the correlation measurements \label{sec:discussg2}}
Using the $g^{(2)}$ measurements introduced above, we first deduce the origin of the \mbox{822.7 nm} luminescence: First, we  emphasize that the different time constants $\tau_1$  and $\tau_1^{NIR}$ exclude that the NIR line is a vibronic sideband: Vibronic sidebands originate from the same excited state as the ZPL; however, they end in vibrationally excited states of the ground state.\cite{Walker1979} These states decay within picoseconds to the vibrational ground state which is also the final state of the ZPL transition.\cite{Lounis2005} This very fast relaxation following the emission of a photon does not influence the measured  $g^{(2)}$ and thus the $g^{(2)}$ function measured for sideband fluorescence is equal to the $g^{(2)}$ function measured for the ZPL fluorescence itself.

\begin{figure}[h]
\includegraphics[width=6.8cm]{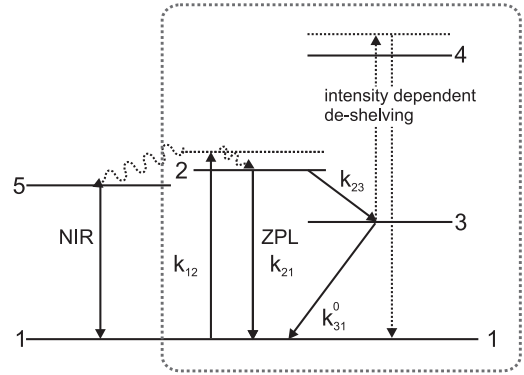}
\caption{Schematic representation of the extended model including the NIR transition. The level scheme included in the dashed box indicates the level scheme for the intensity dependent de-shelving model. Length of the NIR transition and of the ZPL transition drawn to scale. \label{5levelscheme}}
\end{figure}

Second, on the basis of the level scheme depicted in Fig.\ \ref{5levelscheme} (Level 1--4) used to model the ZPL emission dynamics one might suspect that the \mbox{822.7 nm} PL arises from the de-shelving transition $3\to1$. A participation of the ground state is motivated by the observation of absorbing transitions at similar wavelengths in diamonds containing SiV ensembles.\cite{Dhaenens2011} For such a situation, we calculate the $g^{(2)}$ function for the light emitted on the transition $3\to1$ analogously to $g^{(2)}_{ZPL}$
\begin{equation}
g^{(2)}_{3 \to 1}=\frac{N_3(\tau)}{N_3(\tau\rightarrow \infty)},  \label{g231eq}
\end{equation}
as also here each photon detection event projects the system into the ground state and we can use the same initial condition as for the transition $2\to1$. From the discussion of the ZPL emission dynamics, $k_{31}$ was found to be intensity dependent [Eq. (\ref{satdeshrate})]. For our model, we assume that the transition $3\to1$ is radiative. In contrast, the intensity dependent de-shelving process can be interpreted as a non-radiative transition $3\to1$ via state 4. Both processes modify the population of state 3 and thus $g^{(2)}_{3 \to 1}$.  Nevertheless, for each excitation power, the probability for a photon emission on the transition $3\to1$ after a delay $\tau$ is proportional to the population ${N_3(\tau)}$ and thus  Eq. (\ref{g231eq}) indeed describes the $g^{(2)}$ function.

\begin{figure}[h]
\centering
\includegraphics{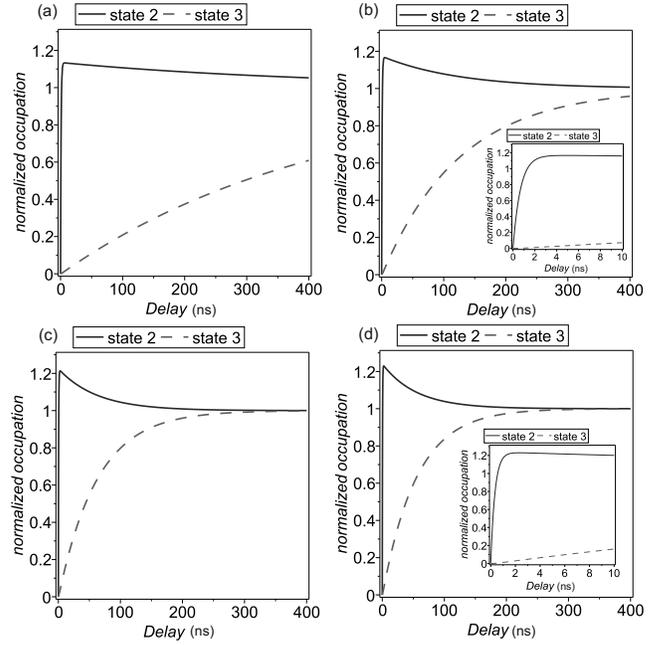}
\caption{Normalized occupation of state 3 and state 2, simulated using $k_{23}=5.8$ MHz, $k_{21}= 1.16$ GHz, $k_{31}^0=0.27$ MHz, $d=18.96$ MHz and $\sigma=8.4$ MHz/$\mu$W and $c=66.6$ $\mu$W. (a) $P=7$ $\mu$W,(b)
$P=35$ $\mu$W, (c) $P=140$ $\mu$W, (d) $P= 210$ $\mu$W. The excitation powers roughly correspond to $0.1\, P_{sat}$, $0.5\, P_{sat}$, $2\, P_{sat}$ and $3\, P_{sat}$. Insets show zoomed images. \label{g2simulations} }
\end{figure}

We now solve the coupled rate equations [Eqs.\ (\ref{ratengl1})--(\ref{ratengl3})] using $k_{21}$, $k_{23}$, $k_{31}^0$, $d$, $c$ and $\sigma$ obtained from the measurements of $g^{(2)}_{ZPL}$.\footnote{Solving of the differential equations (analytically) and calculating the limiting values $N_{2,3}(\tau\rightarrow \infty)$ performed using Maplesoft's Maple 15.} From the solution of the rate equations, we calculate the normalized occupation for states 3 and 2 for different excitation powers $P$. Figure \ref{g2simulations} summarizes the resulting normalized occupation of state 2 ($g^{(2)}_{ZPL}$) as well as of state 3 ($g^{(2)}_{3 \to 1}$), details see figure caption.  Comparing Figs.\ \ref{fig:g2}(a)+(b) and \ref{g2simulations}, it is apparent that the simulated $g^{(2)}$ functions reasonably agree with the measured $g^{(2)}_{ZPL}$ functions, thus the parameters deduced using the intensity dependent de-shelving model appropriately describe the observed SiV center. The simulated functions do not include a correction for the instrument response, thus they display $g^{(2)}(0)=0$, whereas the measured $g^{(2)}$ functions display a non-vanishing value for $g^{(2)}(0)$.

As evident from Fig.\ \ref{g2simulations}, $g^{(2)}_{3 \to 1}$ also displays an antibunching; however, the corresponding time constant $\tau_1^{31}$ is significantly longer than $\tau_1$ or $\tau_1^{NIR}$: Below saturation [Fig.\ \ref{g2simulations}(a) and (b)], $g^{(2)}_{3 \to 1}$ raises to a constant value of one within several \mbox{100 ns} as the rate coefficients $k_{23}$, $k_{31}$ are much smaller than $k_{21}$. Thus, the simulated $g^{(2)}_{3 \to 1}$ is in stark contrast to $g^{(2)}_{NIR}$ measured on the \mbox{822.7 nm} line [see Fig.\ \ref{fig:g2}(c)]. The $g^{(2)}$ measurements thus exclude that the \mbox{822.7 nm} line arises due to a transition between the shelving state (state 3) and the ground state (state 1).
Furthermore, the $g^{(2)}_{NIR}$ measurements exclude that the NIR line arises due to the transition $2\to3$: For a process involving state 2 as an intermediate state,  $\tau_1^{NIR}$  should be at least as long as $\tau_1$.

We emphasize that the ZPL and NIR line cannot arise from a cascaded emission as the cross-correlation measurements do not reveal a bunching, thus we can exclude that one of the transitions populates the excited state of the other transition.\cite{Kiraz2002} Furthermore, the photon energy of the excitation laser is \mbox{1.78 eV} which is well below the sum of the energies for both transitions (\mbox{1.68 eV} and \mbox{1.51 eV}). Additionally, the linear dependence of the fluorescence on excitation power for both transitions excludes multi-photon processes for the excitation.

As the $g^{(2)}_{cross}$ measurements prove that the NIR line and the ZPL originate from the same emitter, we propose an extended level scheme including the NIR transition as depicted in Fig.\ \ref{5levelscheme}. On the basis of the evidence presented above, we assume that the NIR line originates from an independent electronic transition, i.e., a second ZPL. We introduce the NIR transition $5\to1$ as a third relaxation path into the level scheme (Fig.\  \ref{5levelscheme}). As $g^{(2)}_{ZPL}$ matches the $g^{(2)}$ function of a three level system, state 5 should not accumulate population (as it is the case for state 3): This would induce deviations from the three level  $g^{(2)}$ function. The short antibunching time constant $\tau_1^{NIR}$ supports the assumption of a fast decay path. Thus, this pathway, similar to fast non-radiative decay paths, might not lead to a deviation from the $g^{(2)}$ function of a three level system. We assume that the relaxation to state 5 occurs from the intermediate pumping level in the off-resonant pumping process: For a process involving state 2 as an intermediate state,  $\tau_1^{NIR}$  should be at least as long as  $\tau_1$, as the filling of state 5 from state 2 would limit the antibunching time constant. The relaxation rate coefficients from the pumping level are not directly accessible in our measurements. We assume these rate coefficients significantly exceed any other coefficients in the system ensuring that the pumping level does not accumulate population. This assumption is justified, as the vibrationally excited states that are most probably responsible for the pumping transition used here feature very short lifetimes in the picosecond range.\cite{Lounis2005}  The ratio of the relaxation rates from the pumping state to state 5 and to state 2 should influence the branching ratio for the \mbox{740 nm} ZPL and the NIR line. However, taking into account also a potentially differing quantum efficiency, we cannot use the intensity ratio of the lines to deduce the branching ratio.

For a two level system at low excitation powers, $\tau_1$ indicates the lifetime of the excited state of the radiating transition assuming no longer lived states are populated in the excitation process. Thus, from the $g^{(2)}$ measurements, one might expect a very bright emission from the \mbox{822.7 nm} line due to a short lifetime. However, the fraction of the population available for the \mbox{822.7 nm} line is unsettled taking into account the unknown branching ratio of the transition from the pumping level to states 2 and 5. In principle, the lower brightness might be linked to a more efficient quenching mechanism for the NIR luminescence resulting in a low quantum efficiency.\cite{Rogers2010} Non-radiative quenching of low energy transitions has been reported for NV$^-$  centers: The infrared emission at \mbox{1046 nm} is four orders of magnitude weaker than the visible \mbox{637 nm} emission despite the fact that 30\% of the population is available for the infrared transition.\cite{Rogers2008}

In the following, we further analyze the $g^{(2)}_{cross}$ measurements and infer the switching dynamics of ZPL and NIR line. $g^{(2)}_{cross}$ reveals a fully symmetric antibunching. An asymmetric antibunching would indicate that the time elapsing between the emission of an NIR photon and a ZPL photon is different from the time elapsing for the reverse order of events.\cite{Kiraz2002,Regelman2001} In our model, the two lines are populated via the same pumping transition and share the same ground state, consistent with the observation of a symmetric cross-correlation. For the interpretation of the cross-correlation time constant $\tau_1^{cross}$, we consider Ref.\ \onlinecite{Sallen2010} where $g^{(2)}_{cross}$ measurements between disjunct spectral windows of the PL transition of a single quantum dot have been introduced. The emission line is broadened by spectral diffusion, consequently a change of the emission wavelength from one spectral window to the other requires a spectral jump of the emitter. $g^{(2)}_{cross}$ exhibits an antibunching, where $\tau_1^{cross}$ reveals the characteristic time $\tau_d$ for the spectral jumps. Interpreting our measurements analogously, $\tau_1^{cross}$ reveals the characteristic time for the SiV center to change its emission between the NIR and ZPL transitions. As $\tau_1^{cross}$ is comparable to the lifetime of the excited state (state 2), the changes occur in between successive excitation and emission cycles of the ZPL. This short characteristic time also supports the assumption of an alternative decay path connecting the pumping levels and the ground state of the center without involving the long lived shelving state: The longer time windows, in which the emitter resides in the shelving state (state 3) do not influence the measured $g^{(2)}_{cross}$ function between the ZPL and the NIR lines if the level scheme in Fig.\ \ref{5levelscheme} is applicable. Additionally, $g^{(2)}_{cross}$ indicates that the SiV center changes between the two emission lines frequently instead of undergoing a larger number of transitions on either the ZPL or the NIR line before changing its emission wavelength again.
\section{Conclusion}
Photoluminescence (PL) spectra of single SiV centers frequently feature very narrow room temperature PL lines in the near-infrared spectral region, mostly between \mbox{820 nm} and  \mbox{840 nm}.\cite{Neu2011b}  In this work, we clearly prove for a single SiV center, that this PL (at \mbox{822.7 nm}) is due to an additional electronic transition (besides the well known ZPL here at \mbox{739.1 nm}). We find a linewidth of \mbox{1.4 nm} (\mbox{2.6 meV}) for the NIR line. The NIR line saturates at similar excitation power as the ZPL. However, despite a shorter excited state lifetime deduced from the $g^{(2)}$ measurements, it delivers a factor of four lower fluorescence intensity. A $g^{(2)}_{cross}$ measurement between the ZPL  and the NIR line reveals anti-correlated emission, thus proving that the emission originates from the same emitter. The short antibunching time constant in the $g^{(2)}_{cross}$ measurements indicates a fast switching between the transitions. $g^{(2)}$ auto-correlation measurements exclude that the NIR emission arises due to a transition from/to the shelving state of the SiV center. They further exclude that the NIR line is a vibronic sideband. Polarization measurements reveal preferential absorption of linearly polarized light for both transitions. The maximum absorption is observed for the same polarization direction. The emitted fluorescence light shares a linear polarization direction. We interpret the NIR transition as an independent electronic transition populated from the same pumping levels as the ZPL transition.
\begin{acknowledgments}
We acknowledge funding by the DFG  and the BMBF (EPHQUAM 01BL0903).
\end{acknowledgments}
%

\end{document}